\documentclass[
 reprint,
 superscriptaddress,
 amsmath,
 amssymb,
 aps,
 prb
]{revtex4-2}

\usepackage{graphicx}
\usepackage{dcolumn}
\usepackage{bm}
\usepackage[hidelinks]{hyperref}

\hypersetup{
  colorlinks   = true,
  urlcolor   = blue,
  citecolor   = blue
}

\begin{document}


\title{Van Hove singularities at the $L$-face of the lutetium nitride phonon dispersion}

\author{M.~Markwitz}\email{martin.markwitz@vuw.ac.nz}
 \affiliation{Robinson Research Institute, Victoria University of Wellington, P.O. Box 33436, Petone 5010, New Zealand}
\author{K.~Kneisel}
 \affiliation{The MacDiarmid Institute for Advanced Materials and Nanotechnology, Victoria University of Wellington, P.O. Box 600, Wellington 6140, New Zealand}
 \affiliation{School of Chemical and Physical Sciences, Victoria University of Wellington, P.O. Box 600, Wellington 6140, New Zealand}
\author{R.~G.~Buckley}
 \affiliation{Robinson Research Institute, Victoria University of Wellington, P.O. Box 33436, Petone 5010, New Zealand}
 \affiliation{The MacDiarmid Institute for Advanced Materials and Nanotechnology, Victoria University of Wellington, P.O. Box 600, Wellington 6140, New Zealand}
\author{K.~C.~Rule}
 \affiliation{School of Physics, Faculty of Engineering and Information Science, University of Wollongong, Northfields Avenue, Wollongong, New South Wales 2522, Australia}
 \affiliation{Australian Centre for Neutron Scattering, Australian Nuclear Science and Technology Organisation, Lucas Heights, New South Wales 2234, Australia}
\author{M.~Chegeni}
 \affiliation{The MacDiarmid Institute for Advanced Materials and Nanotechnology, Victoria University of Wellington, P.O. Box 600, Wellington 6140, New Zealand}
 \affiliation{School of Chemical and Physical Sciences, Victoria University of Wellington, P.O. Box 600, Wellington 6140, New Zealand}
\author{H.~J.~Trodahl}
 \affiliation{School of Chemical and Physical Sciences, Victoria University of Wellington, P.O. Box 600, Wellington 6140, New Zealand}
\author{W.~F.~Holmes-Hewett}
 \affiliation{Robinson Research Institute, Victoria University of Wellington, P.O. Box 33436, Petone 5010, New Zealand}
 \affiliation{The MacDiarmid Institute for Advanced Materials and Nanotechnology, Victoria University of Wellington, P.O. Box 600, Wellington 6140, New Zealand}
\author{J.~D.~Miller}
 \affiliation{Robinson Research Institute, Victoria University of Wellington, P.O. Box 33436, Petone 5010, New Zealand}
 \affiliation{The MacDiarmid Institute for Advanced Materials and Nanotechnology, Victoria University of Wellington, P.O. Box 600, Wellington 6140, New Zealand}
\author{F.~Natali}
 \affiliation{The MacDiarmid Institute for Advanced Materials and Nanotechnology, Victoria University of Wellington, P.O. Box 600, Wellington 6140, New Zealand}
 \affiliation{School of Chemical and Physical Sciences, Victoria University of Wellington, P.O. Box 600, Wellington 6140, New Zealand}
 \affiliation{Liquium Ltd., 69 Gracefield Road, Gracefield, Lower Hutt 5010, New Zealand}
\author{M.~Maddah}
 \affiliation{Liquium Ltd., 69 Gracefield Road, Gracefield, Lower Hutt 5010, New Zealand}
\author{B.~J.~Ruck}
 \affiliation{The MacDiarmid Institute for Advanced Materials and Nanotechnology, Victoria University of Wellington, P.O. Box 600, Wellington 6140, New Zealand}
 \affiliation{School of Chemical and Physical Sciences, Victoria University of Wellington, P.O. Box 600, Wellington 6140, New Zealand}
\author{S.~Granville}
 \affiliation{Robinson Research Institute, Victoria University of Wellington, P.O. Box 33436, Petone 5010, New Zealand}
 \affiliation{The MacDiarmid Institute for Advanced Materials and Nanotechnology, Victoria University of Wellington, P.O. Box 600, Wellington 6140, New Zealand}

\date{\today}

\begin{abstract}
We report the structural and vibrational properties of the prototypical 4$f$-filled nonmagnetic member LuN of the lanthanide nitrides, \textit{Ln}N, with elastic and inelastic neutron scattering data at $4$~K. We find a peak in the generalized density of states which, through input from a DFT+$U$ computation, we ascribe to a van Hove singularity on the fourfold-degenerate $L$-face of the Brillouin zone. This work advances the understanding of phonon dynamics in \textit{Ln}N beyond the $\Gamma$-point.
\end{abstract}

\maketitle

\section{Introduction}

The lanthanide nitrides (\textit{Ln}N) are a group of rock-salt structured semiconducting materials with a variety of magnetic properties at cryogenic temperatures~\cite{Natali2013,holmes2025rare}. These materials may find application in on-chip cryogenic electronics or in ammonia production due to the strong independent tunability of their magnetic and electric properties, and their capacity for nitrogen cracking at ambient temperature and pressure~\cite{xiao1996proximity,Senapati2011,Pal2013,Pal2014,Cascales2019,devese2022non,Pot2023,miller2025complete,Ullstad2019,chan2020facile,chan2023}. Their electronic, optical, and magnetic properties were researched in the 1960's-1970's, then largely forgotten, and with advances in sample preparation and characterization, research has restarted since 2005. Experimental investigations of their vibrational properties, however, have to date almost entirely been restricted to zone centre vibrations.

The vibrational and thermal properties of this class of materials are difficult to study due to the materials' propensity toward degradation in ambient conditions~\cite{Kneisel2024,uhlemann2026ambient}. This, in conjunction with an initially inappropriate density functional theory (DFT) computational description has confused the initially determined properties of these materials, suggesting a metallic ground state for many of the \textit{Ln}N~\cite{petukhov1996electronic,duan2005strain,aerts2004half}. Recent investigations of both nominally stoichiometric experimental and theoretical works have generally established insulating ground states for the \textit{Ln}\textsuperscript{3+}N\textsuperscript{3-} series~\cite{larson2007electronic,topsakal2014accurate,galler2022electronic,trodahl2007ferromagnetic,warring2014ybn,holmes20194}. Perhaps driving the historical confusion, the \textit{Ln}N can also be prepared in \textit{nitrogen-poor} conditions, leading to large concentrations of nitrogen vacancies, driving metallic-like $n$-type conductivity~\cite{Punya2011,Ruck2012,holmes2020nitrogen,holmes2021electronic,devese2022probing,holmes2024spin,markwitz2026optolectronic}. 

Fourier-transform infrared spectroscopy (FTIR) is the go-to method for studying IR-active parts of the vibrational spectrum in rock salt crystals, and while it is only sensitive to frequencies at the $\Gamma$ point, it allows measurement of the TO($\Gamma$) mode in the \textit{Ln}N. The TO($\Gamma$) mode in LuN has been reported to be at $273$~cm\textsuperscript{-1} ($34$~meV)~\cite{holmes2022gamma}. Other methods such as infrared spectroscopic ellipsometry can also provide information about the TO($\Gamma$) mode, as recently demonstrated in ScN~\cite{grumbel2024band}. Although Raman spectroscopy is often a practical technique used for accessing long-range vibrational modes complementary to FTIR, in rock salt structures the Raman selection rules prohibit the excitation of the TO($\Gamma$) and LO($\Gamma$) modes~\cite{merlin1978multiphonon}. Surprisingly, in the rock salt-structured \textit{Ln}N and YbX/EuX (X$=$O, S, Se, Te), strong Raman signals were observed~\cite{Granville2009,VanKoughnet2023,Kneisel2024}. These phonons generally appear with energies between $550$ and $700$~cm\textsuperscript{-1} ($70$ to $85$~meV), in particular, the signal appears in LuN at $587$~cm\textsuperscript{-1}. The Fröhlich interaction enables this; alleviating of the Raman selection rules through a displacement of the zero-point vibration amplitude of LO phonons and a dispersionless 1$s$ exciton~\cite{merlin1978multiphonon}. As mentioned earlier, beyond-zone center vibrations in the \textit{Ln}N have been scarcely investigated, which may possess van Hove singularities that can significantly influence the thermal properties of these compounds, and notably change the thermal conductivity and heat capacity at specific temperatures.

To probe vibrations beyond the zone centre inelastic neutron scattering (INS) is in general employed. INS provides a direct measure of the (weighted) phonon density of states, free of the optical transition selection rules which can then be compared to optical measurement techniques and DFT calculations. So far, only a few reports of INS exist on the \textit{Ln}N, studying primarily crystal field states and transitions using triple-axis spectrometers (TAS) to probe NdN and YbN~\cite{warming1975crystal,furrer1976crystal,donni1990fcc}, or time-of-flight methods to study PrN, TmN, and HoN~\cite{mook1972crystal,davis1973direct,davis1975neutron}. Measurements of the crystal field transitions in the \textit{Ln}N as a whole necessitate reference measurements of a nonmagnetic nitride, LuN, that does not exhibit crystal field transitions due to its filled 4$f$ shell. Measurements of the phononic response can be directly compared to LuN for the rest of the \textit{Ln}N due to their otherwise similar vibrational properties. The half-filled 4$f^{7}$ shell of GdN would also be interesting due to the lack of crystal field states, but are impractical due to the large thermal neutron absorption cross section of Gd.

The aim of this work is to characterize the phononic properties of LuN by measuring its phonon density of states with INS. We characterize our powders using conventional techniques (XRD, Raman), and analyze the structural and vibrational properties of the LuN powder using neutron scattering. Accordingly, with support from DFT calculations, we identify a peak in the measured phonon density of states (PDOS) as a van Hove singularity of the lower branch of optical modes on the $L$-face of the Brillouin zone.


\section{Results}

\begin{figure}
    \centering
    \includegraphics[width=\linewidth]{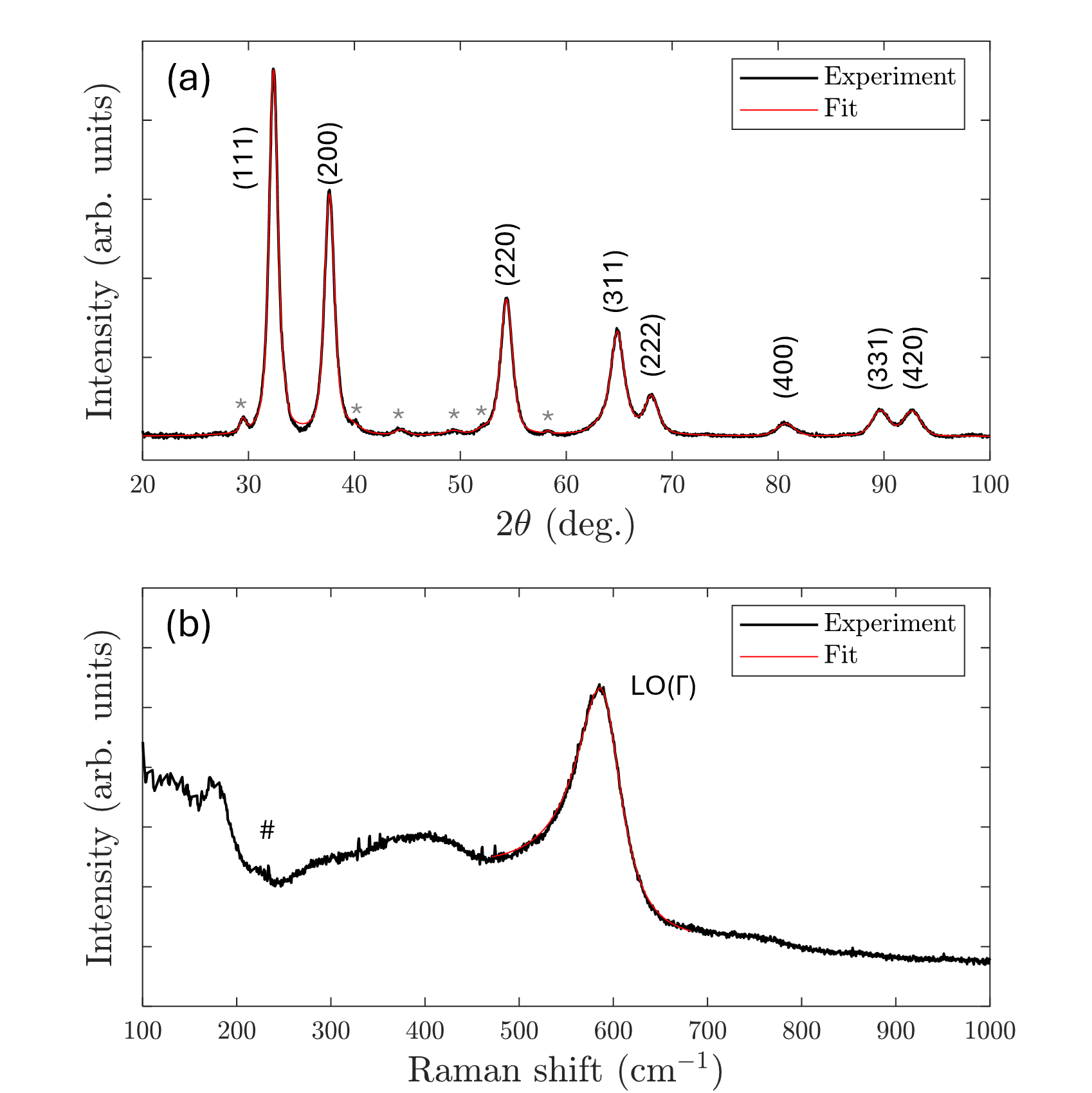}
    \caption{Ambient (a) XRD ($\lambda_{\mathrm{Cu~K}\alpha_{1}}=1.540594$~{\AA}) and (b) Raman measurements ($\lambda=514.5$~nm) of LuN powder. The asterisks in the XRD pattern correspond to Lu metal-derived signal. The hash in the Raman signal at $\approx210$~cm\textsuperscript{-1} corresponds to a defect-activated Raman mode. The experimental data are the black lines, and the fits to the data are red.}
    \label{fig:LuN_basicCharacterization}
\end{figure}

The LuN powder preparation process is discussed in the methods section. Ambient temperature XRD measurements of LuN powders are presented in Figure~\ref{fig:LuN_basicCharacterization}(a). The dominant peaks as labeled in the figure correspond to LuN. We note that there are small signals originating from Lu which has not reacted with nitrogen during fabrication. The XRD pattern was fitted using pseudo-Voigt functions, which identified a lattice constant of $4.7648(3)$~{\AA}. This is in agreement with previous reported lattice constants on LuN thin films and powders~\cite{devese2022probing,Kneisel2024,rivera2026exchange,su2024synthesis}. Figure~\ref{fig:LuN_basicCharacterization}(b) shows a strong Raman signal corresponding to the conventionally-forbidden LO($\Gamma$) mode pointing to the breakdown of the usual Raman selection rules in LuN, as also found in the analogous-structured YbX and EuX~\cite{merlin1978multiphonon,VanKoughnet2023}. Using a Fano lineshape we find a LO($\Gamma$) mode frequency of $587(1)$~cm\textsuperscript{-1} ($72.8(1)$~meV) which is also in good agreement with previous reports on the thin film and powder forms of LuN~\cite{VanKoughnet2023,Kneisel2024,markwitz2025raman}. Beyond this, a clear multiphonon Raman signal is observed at Raman shifts of $\approx175$~cm\textsuperscript{-1}, matching the structure of data which was acquired for thin film LuN prepared in the presence of nitrogen ions~\cite{VanKoughnet2023}. Finally, a small vibrational mode marked with a \# at $\approx225$~cm\textsuperscript{-1} suggests that a Raman active point defect structure exists within the powder~\cite{VanKoughnet2023}.

\begin{figure}
    \centering
    \includegraphics[width=\linewidth]{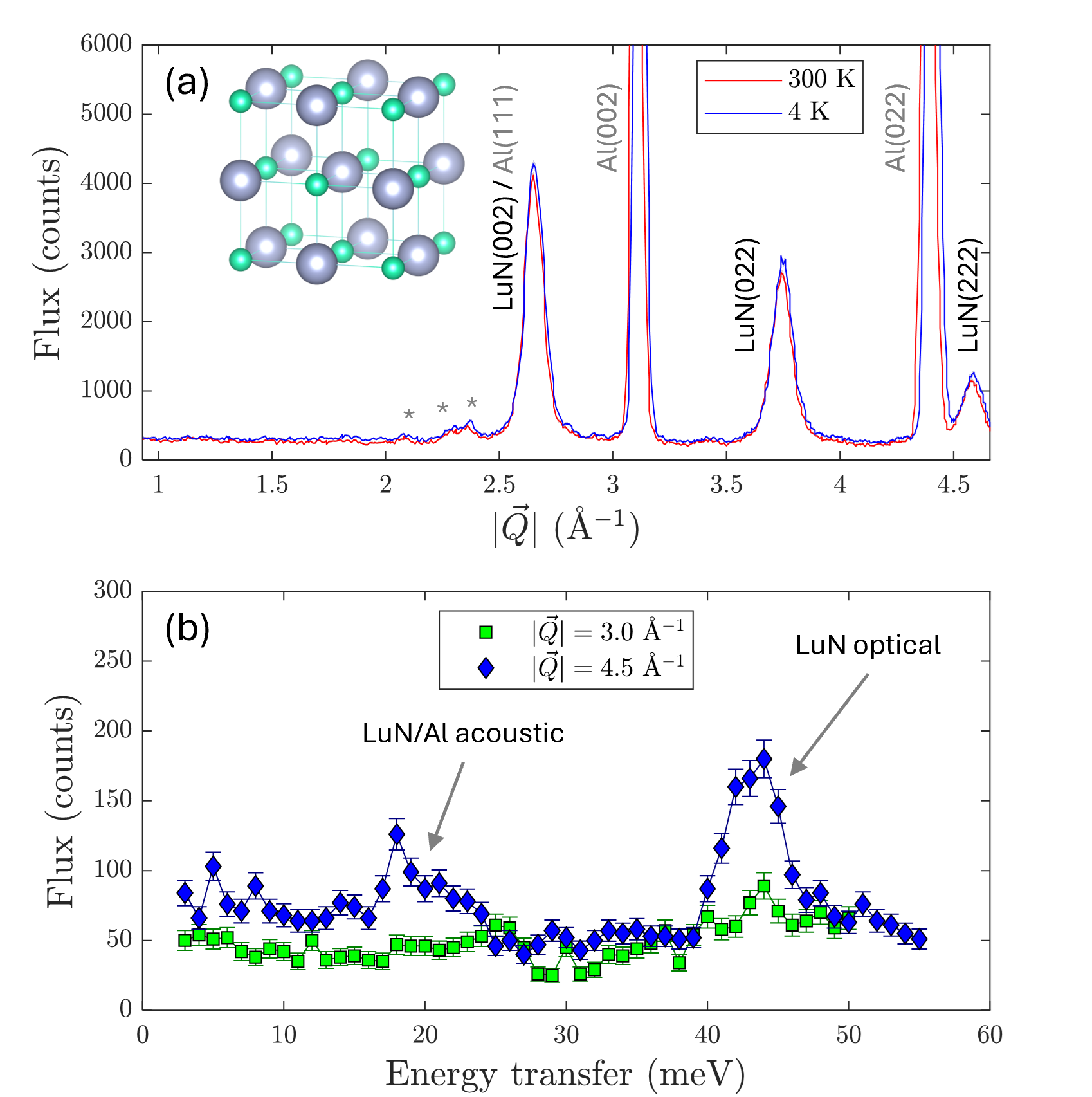}
    \caption{(a) Neutron diffraction patterns from LuN powder ($\lambda_{\text{n}}=2.345$~{\AA}) at $300$~K (red) and $4$~K (blue). The asterisks in the neutron diffraction pattern correspond to Lu metal-derived signal. The inset graphic is the rock salt crystal structure of LuN, (Lu atoms are green and N atoms are grey). (b) INS spectra with $|\vec{Q}|=3.0$~{\AA}\textsuperscript{-1} (green) and $|\vec{Q}|=4.5$~{\AA}\textsuperscript{-1} (blue) for LuN powder at $4$~K.}
    \label{fig:LuN_neutron}
\end{figure}

Figure~\ref{fig:LuN_neutron}(a) shows the neutron diffraction measurements conducted using the thermal triple axis spectrometer, Taipan at the Australian Nuclear Science and Technology Organisation (ANSTO). Similarly to the XRD data, this data identifies LuN and Lu phases present in the LuN powder sample, along with Al originating from the sample holder and cryostat. The absence of the LuN(111) signal is due to the similar thermal neutron scattering cross sections of Lu ($6.53$~barn) and N ($11.51$~barn), while the low intensity of the Al(111) is a result of the sample holders' crystalline orientation. At room temperature the neutron diffraction measurement resulted in a lattice constant at $300$~K of $4.750(2)$~{\AA}. The discrepancy in lattice constant measured between the neutron diffraction and XRD measurements is likely due to slight differences in goniometer calibration and sample age. After cooling the sample to $4$~K the lattice constant of the LuN-associated peaks contracted to $4.743(2)$~{\AA}, leading to a percentage change of $-0.14(7)$~\%, or an absolute contraction of $-0.007(4)$~{\AA} compared to the measurement at $300$~K. While differing from one another, the extracted $d$-spacings from the neutron scattering data and XRD data individually support an FCC crystal structure at both temperatures, drawn as the inset in Figure~\ref{fig:LuN_neutron}(a). Finally, no additional peaks or diffraction line intensifications are observed at low $|\vec{Q}|$ when comparing the $4$~K data to the $300$~K data in the LuN neutron diffraction measurement, ruling out any onset of a significant contaminant ferromagnetic or antiferromagnetic phase in the present sample.

Immediately following the neutron diffraction measurement, INS measurements were conducted and are depicted in Figure~\ref{fig:LuN_neutron}(b) which highlight phonon-derived signals from LuN. Two measurements were conducted, one with a low momentum transfer (green squares, $|\vec{Q}|=3.0$~{\AA}\textsuperscript{-1}), and high momentum transfer (blue diamonds, $|\vec{Q}|=4.5$~{\AA}\textsuperscript{-1}). A peak at an energy transfer of $43(1)$~meV in the $|\vec{Q}|=4.5$~{\AA}\textsuperscript{-1} measurement and reduction in the $|\vec{Q}|=3.0$~{\AA}\textsuperscript{-1} measurement signals phononic origin of the signal, as the scattering contribution scales in intensity by the area of the sampled $|\vec{Q}|$-sphere, i.e., $\propto|\vec{Q}|^{2}$. Furthermore at low energy transfers a phononic signal between $16$ and $25$~meV is found which can be attributed to the density of states of either the LuN or Al acoustic branches~\cite{tang2010anharmonicity}. Since there is no energy transfer at which there is a signal excess in the $|\vec{Q}|=3.0$~{\AA}\textsuperscript{-1} data compared to the $|\vec{Q}|=4.5$~{\AA}\textsuperscript{-1}, we can exclude magnons or crystal field transitions occurring in our LuN sample in the measured energy range.

\section{Discussion}

\begin{figure}
    \centering
    \includegraphics[width=\linewidth]{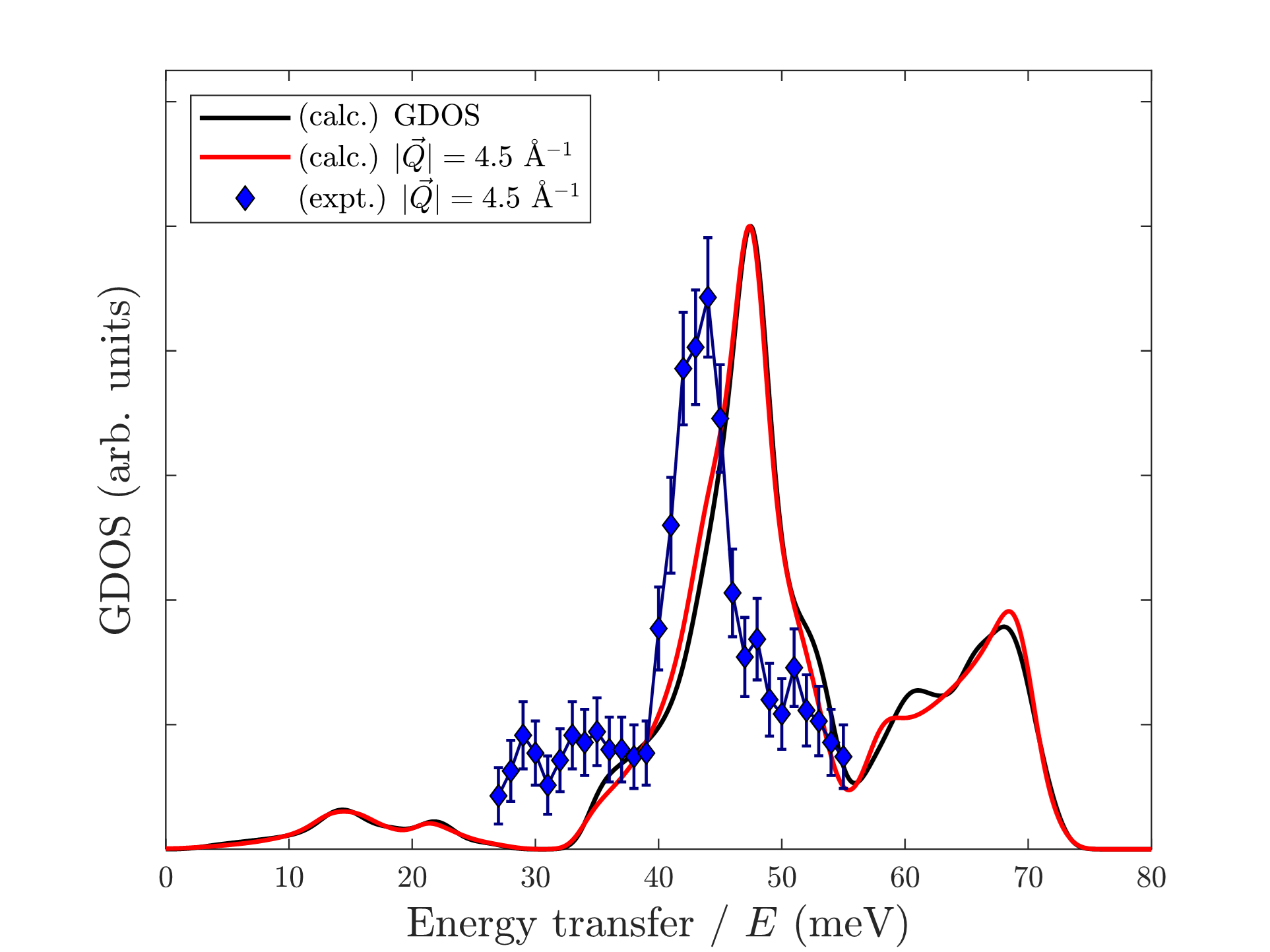}
    \caption{LuN DFT+$U$ calculated plotted against the INS experimental generalized density of states. The black line is the true GDOS, the red line is the GDOS sampled from a $|\vec{Q}|=4.5$~{\AA}\textsuperscript{-1} surface, and the blue diamonds are the same $|\vec{Q}|=4.5$~{\AA}\textsuperscript{-1} INS data as in Figure~\ref{fig:LuN_neutron}(b).}
    \label{fig:LuN_DFT}
\end{figure}

\begin{figure*}
    \centering
    \includegraphics[width=\linewidth]{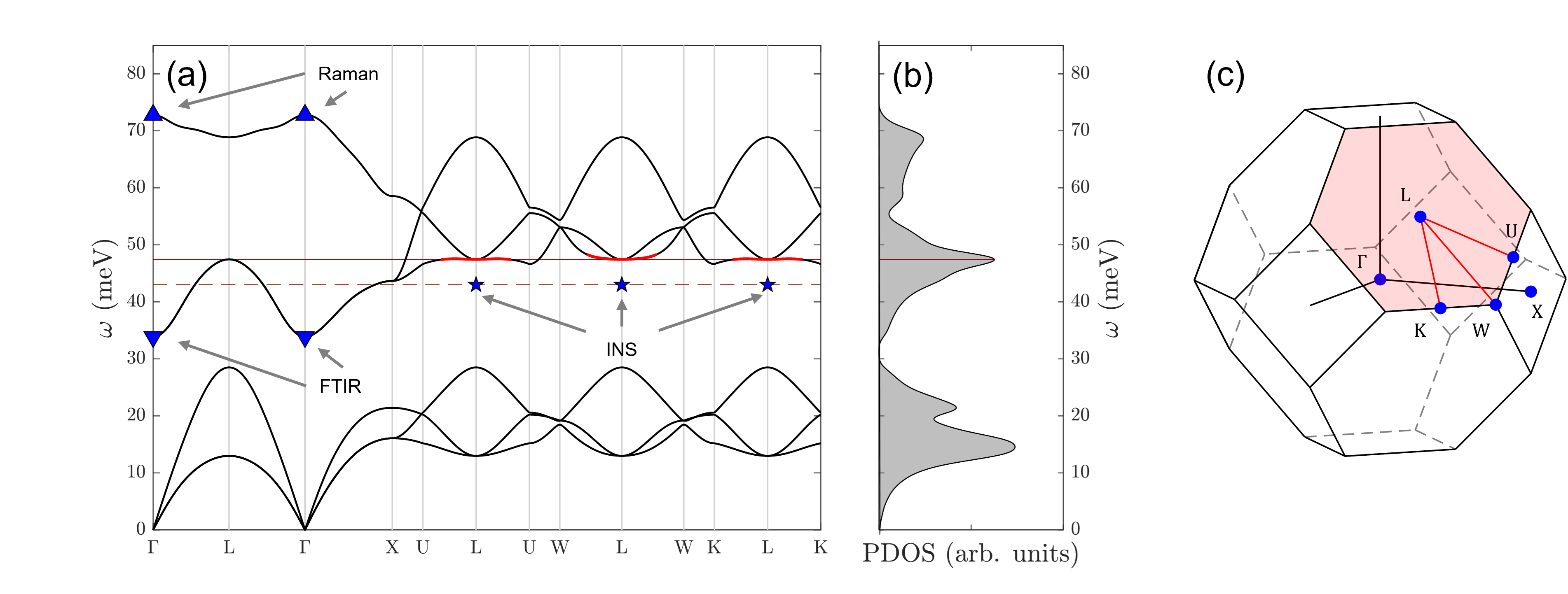}
    \caption{(a) Calculated phonon dispersion for LuN using DFPT+$U$ (black lines) along high symmetry paths to highlight the lack of dispersion on the $L$ face of the first Brillouin zone. The peak in the calculated (experimental) GDOS is drawn as the filled (dashed) red line at $43$~meV ($47.4$~meV). The FTIR, Raman, and INS-derived TO($\Gamma$), LO($\Gamma$), and optical $L$-face van Hove singularity signals are included as the blue inverted triangles, blue upright triangles, and blue star symbols. The lowest energy optical mode contributing to the GDOS signal on the $L$ face is highlighted as the thick red lines. The peak in the calculated GDOS coincides with a van Hove singularity at the $L$-face, drawn as the dotted red line. (b) LuN PDOS computed using a uniform $\mathbf{q}$ grid. (c) First Brillouin zone of LuN with a highlighted $L$ face (red) and annotated high symmetry points (blue).}
    \label{fig:LuN_dispersion}
\end{figure*}


The LuN phonon signal from inelastic neutron scattering at an energy transfer of $\approx43$~meV lies between the measured TO($\Gamma$) and LO($\Gamma$) mode frequencies from FTIR and Raman measurements, respectively. We compare the experimental results to those computed \textit{ab initio} using DFT calculations. Figure~\ref{fig:LuN_DFT} plots the calculated generalized density of states (GDOS) drawn using a black line and compares it to the \textit{effective}-GDOS ($e$-GDOS) drawn using a red line calculated from a $|\vec{Q}|=4.5$~{\AA}\textsuperscript{-1} sphere, suggesting that the $|\vec{Q}|=4.5$~{\AA}\textsuperscript{-1} measurement resembles the GDOS. Notably, the acoustic vibrations are primarily associated with Lu atomic vibrations, giving them a weak GDOS due to the large mass of Lu, while the strongest contribution comes from the optical modes, which are primarily associated with N atomic vibrations. Outside some highly-dispersive high energy modes between $55$ and $75$~meV, the $|\vec{Q}|=4.5$~{\AA}\textsuperscript{-1} sphere well-approximates the shape of the true GDOS, thereby giving confidence about the $|\vec{Q}|=4.5$~{\AA}\textsuperscript{-1} momentum transfer measurement drawn using blue diamonds being a representative measurement of the true LuN GDOS. The peak in the GDOS and $e$-GDOS is at $47.4$~meV, which is $\approx4$~meV above the experimental peak. The difference in energy of these peaks is likely a result of the high-$\mathbf{q}$ force constants calculated from density functional perturbation theory+$U$ (DFPT+$U$), and does not consider bond anharmonicity or any self-energy corrections, the study of which are outside the scope of this work. Lastly, we the significant signal excess below $25$~meV in the experiment shown in Figure~\ref{fig:LuN_neutron}(b) compared to the GDOS computation shown in Figure Figure~\ref{fig:LuN_DFT} indicates a significant contribution from the Al sample container in the measured INS signal. Given that the greatest crystalline contaminant in the LuN powder is Lu metal, and Lu having a low thermal neutron scattering cross section and high atomic mass, negligible contribution to this low energy transfer signal originates from any in-powder contamination. Given the agreement between the computed $e$-GDOS and the $4.5$~{\AA}\textsuperscript{-1} measurement we are confident that the computed phonon-dispersion accurately represents the true material, within the $4$~meV disagreement.

Using a phonon dispersion, the $\mathbf{q}$ space origin of the peak in the GDOS can be studied. Figure~\ref{fig:LuN_dispersion}(a) shows the calculated phonon dispersion along high symmetry lines throughout the Brillouin zone (black lines) and the experimental data points measured from FTIR and Raman spectroscopy as the blue inverted triangles and blue upright triangles. The peak in the calculated GDOS (horizontal red line) coincides with the flat-band optical phonon dispersion at the center of the fourfold symmetric $L$ face of the first Brillouin zone while the experimental signal peaks $4$~meV below it. The lowest-energy optical phonon clearly shows a van Hove singularity on the $L$ face of the first Brillouin zone, highlighted as a wide red line. Normal to the face (toward $\Gamma$) this mode exhibits high dispersion towards the TO($\Gamma$) mode. The phonon density of states (PDOS) without neutron scattering or thermal weighting is drawn in Figure~\ref{fig:LuN_dispersion}(b). We include the first Brillouin zone of FCC crystals as Figure~\ref{fig:LuN_dispersion}(c) where we highlight the $L$ face of the first Brillouin zone in red, highlight the $ULU$, $WLW$, and $KLK$ high symmetry lines, and annotate the high-symmetry points with blue dots. This figure clearly demonstrates in conjunction with panels (a) and (b) that the peak in the PDOS of the optical branch originates primarily from the van Hove singularity on the $L$ face. While the calculated dispersion and PDOS of the acoustic modes also show a van Hove singularity on the $L$ face, which matches the peak in the signal at lower energies measured in Figure~\ref{fig:LuN_neutron}(b), the signal may also originate from the Al sample holder.

This suggests that while the force constants for DFPT+$U$ can be adjusted using experimental inputs to generate experimental values near $\Gamma$, the agreement with experiment does not propagate toward the zone boundaries. This is likely a limit of the exchange correlation potential approximation and the derived short-range force constants. Regardless of the exact agreement with the theoretical calculation, we can deduce that the large area van Hove singularity of the degenerate optical phonon dispersion near the center of the $L$ face is the origin of the strong signal in the INS measurement, proving the first clear investigation of off-$\Gamma$ measurements of phonons in the \text{Ln}N. Using a TAS and interpretation of the experimental data in terms of the $|\vec{Q}|$-sphere can extend also to the rest of the \textit{Ln}N studied using a TAS~\cite{warming1975crystal,furrer1976crystal,donni1990fcc}. Those works studied the the inelastic neutron scattering spectra of a variety of \textit{Ln}N and assigned signals in these data to crystal field transitions or phonons depending on their evolution with $|\vec{Q}|$ or temperature. The assignments of signals measured so far are somewhat sporadic with no consistent observation or assignment to optical phonons, and assignment of features between $10-20$~meV which may originate from sample holders was not considered. Further INS measurements to higher energy transfers are desirable on \textit{Ln}N samples on a variety of the nitrides which can expose the upper branch of optical modes. Now that stoichiometric powders can be readily prepared, measurements which study the crystal field transitions in a variety of \textit{Ln}N should be taken with respect to nonmagnetic analogous crystals, including that provided here for LuN to refine the derived crystal field parameters. 

The effects of van Hove singularities in the phonon dispersion manifest in sudden increases of the heat capacity as a function of temperature and therefore also in the thermal conductivity, which is as-of-yet rarely studied in this class of compounds. Developing the knowledge of the thermal transport properties of these compounds can help provide certainty regarding their application in thermoelectric energy conversion, where efficiency relies on the complex interplay of carrier mobility, carrier density, and thermal conductivity. This research interest has seen recent study within heavier nitride (ErN and YbN) thin films already~\cite{upadhya2021high,Upadhya2022,loyal2023coexistence} in which the heavier cation in those compounds may lead to enhanced energy harvesting performance thanks to a reduced thermal conductivity compared to the more-understood but well-performing ScN in epitaxial thin film form~\cite{kerdsongpanya2011anomalously,burmistrova2013thermoelectric}. Finally, it may be possible to tune the energy of the van Hove singularity and therefore the thermal transport properties by fabricating lanthanide solid solutions (\textit{Ln},\textit{Ln}')N. Such work has found applications in cryogenic electronics in the context of mixed magnetic behavior and band structures of (Gd,Dy)N, (Gd,Sm)N, (Gd,Lu)N, and (Nd,Dy)N~\cite{pot2021contrasting,Pot2023,porat2024tuneable,miller2025complete,rivera2026exchange,vankoughnet2026mixed}. On the other hand, it may also be possible to further enhance the cryocooling capabilities of the nitrides in general by tuning their magnetic and thermal properties through alloying with other \textit{Ln}~\cite{nakagawa2004magnetocaloric,nakano2012ern}.

While powder INS can provide useful information about the optical modes and crystal field excitations in the \textit{Ln}N, it is much less sensitive to acoustic phonons. Inelastic x-ray scattering (IXS) can be used to investigate these otherwise hard-to-access modes in the \textit{Ln}N without requiring large volumes of material. This makes measurements on epitaxial thin films possible, such as those already demonstrated for EuO and ScN~\cite{pradip2016lattice,uchiyama2018phonon}. These complementary techniques can be used to probe the connections between the structure, magnetism, and electronic states through interactions with phonons in the \textit{Ln}N. 

\section{Conclusion}

We have probed the generalized phonon density of states of LuN powders by combining $4$~K inelastic neutron scattering and DFT calculations. We find that the measured INS spectra away from the Brillouin zone centre modes are consistent with a van Hove singularity at the $L$ face of the Brillouin zone as observed in the DFT calculation. When combined with earlier IR and Raman measurements we have an extended agreement between experiment and calculation and thus have significantly advanced our knowledge of the phonon band structure of LuN. This work sets the foundation to study the vibronic and magnonic properties in the (mostly) ferromagnetic insulating \textit{Ln}N using inelastic neutron scattering, benefitting the understanding of this class of understudied materials as a whole.

\section{Acknowledgments}

This research was supported by the New Zealand Ministry of Business, Innovation and Employment (Grant No.~RSCHTRUSTVIC2447). Neutron beam time was awarded at ANSTO under Proposal No. P21184. The MacDiarmid Institute is supported under the New Zealand Centres of Research Excellence programme. William Holmes-Hewett is supported by the Royal Society under the New Zealand Mana T\={u}\={a}papa Future Leader Fellowship (Contract No.~MTP-VUW2402). The computations were performed on R$\overline{\text{a}}$poi, the high performance computing facility of Victoria University of Wellington.

\appendix
\section{Methods}

\subsection{Theoretical methods}

\begin{figure}
    \centering
    \includegraphics[width=\linewidth]{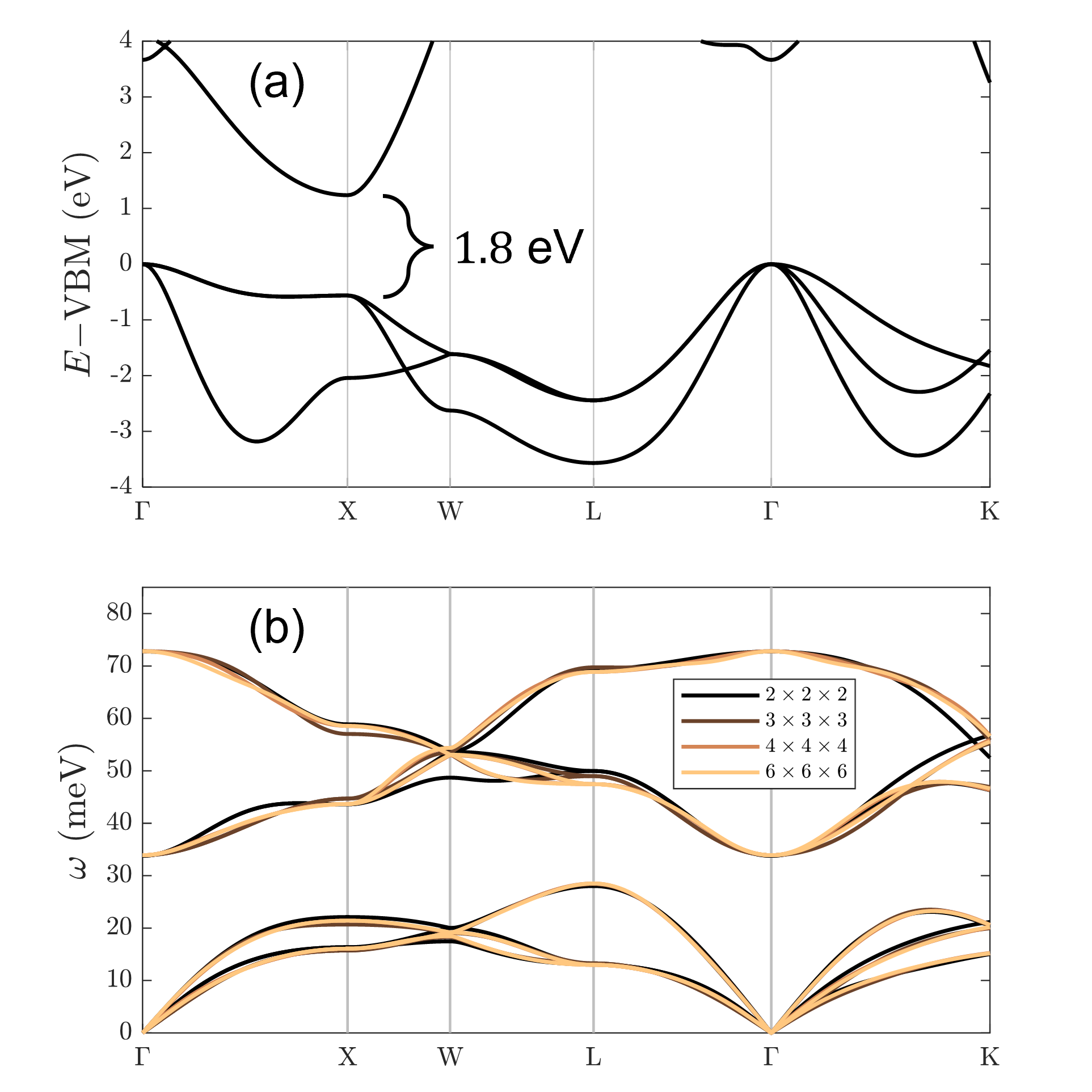}
    \caption{(a) LuN DFT+$U$ band structure calculated using $U=5.3$~eV. (b) LuN DFPT+$U$ phonon dispersion convergence testing for a variety of $\mathbf{q}$ mesh grids using $U=5.3$~eV.}
    \label{fig:LuN_convergence}
\end{figure}

\begin{figure*}
    \centering
    \includegraphics[width=\linewidth]{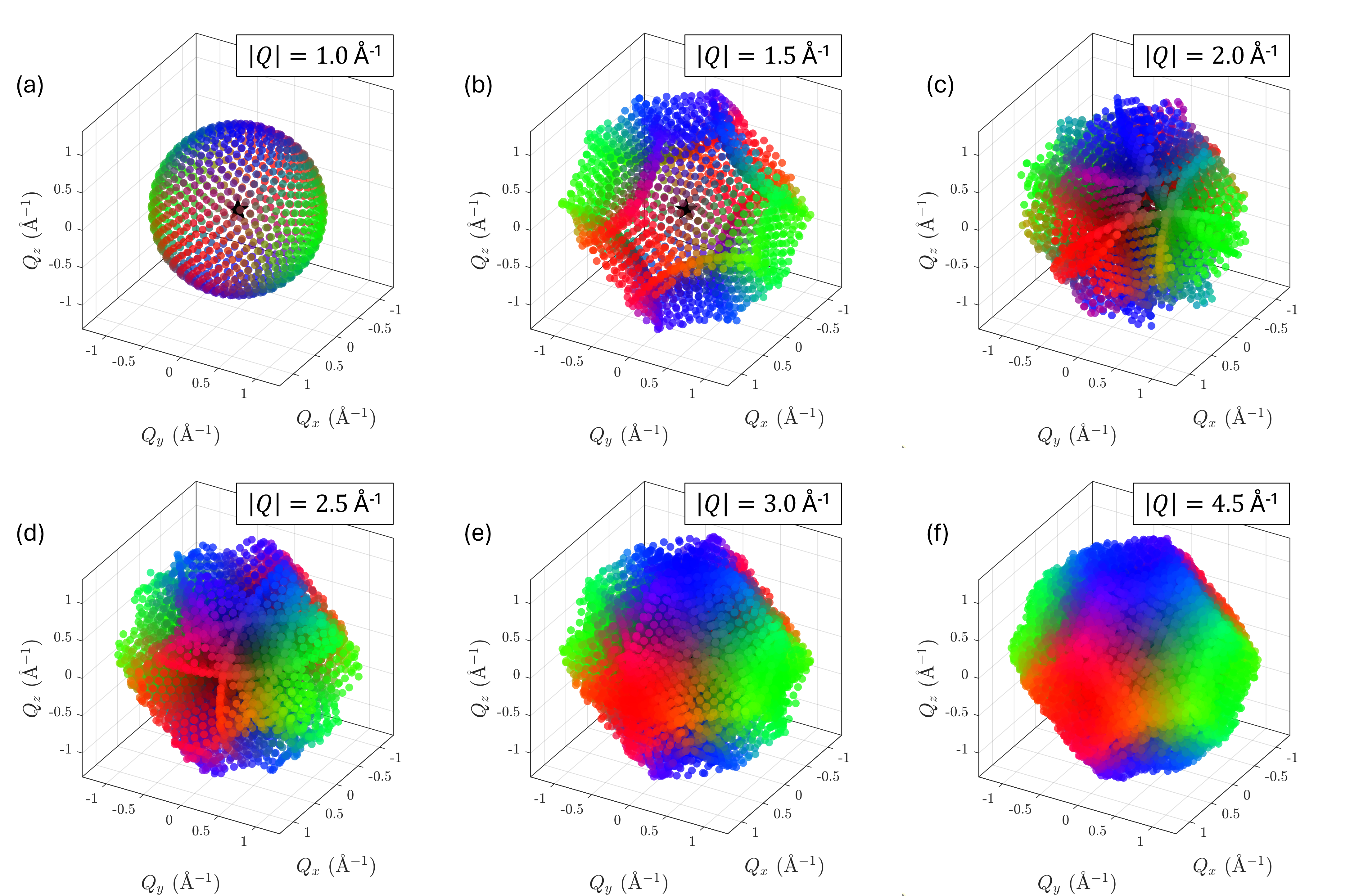}
    \caption{Set of $\mathbf{q}$ points used in our DFT calculations for computing the equivalent set of momentum transfers $|Q|$ measuring using INS. The calculated set of points in each panel (a)-(f) are folded into the first Brillouin zone. The star denotes the origin of the Brillouin zone.}
    \label{fig:Q}
\end{figure*}

We conducted DFT computations using \texttt{Quantum~ESPRESSO} with ultrasoft pseudopotentials from the Dal Corso library for both Lu and N, placing the Lu 4$f$ states in the core~\cite{giannozzi2009quantum,giannozzi2017advanced,giannozzi2020quantum,dal2014pseudopotentials}. The charge density for LuN in the Fm$\overline{\text{3}}$m rock salt crystal structure was converged using a $12\times12\times12$ $\mathbf{k}$ point grid and a wavefunction energy cutoff of $60$~Ry ($816$~eV), which uses the experimental lattice constant derived from the XRD measurement. DFT+$U$ and DFPT+$U$ with a rotationally-symmetric $U\equiv U_{d,\text{eff}}=U_{d}-J_{d}=5.3$~eV with $J_{d}=0$~eV is used to describe the $d$ manifold to ensure the TO($\Gamma$) for LuN matches its experimental value of $273$~cm\textsuperscript{-1} ($34$~meV)~\cite{dudarev1998electron,floris2011vibrational,floris2020hubbard}. The derived direct gap at the $X$ point is $1.8$~eV which slightly overestimates the experimentally-estimated value of $1.7$~eV~\cite{devese2022non}. The calculated band structure for LuN is drawn in Figure~\ref{fig:LuN_convergence}(a).

Given that an estimate for the high frequency permittivity of $\varepsilon_{\infty}=6.29\varepsilon_{0}$ has previously been provided by Xue~\textit{et~al.}~\cite{xue2000dielectric}, where $\varepsilon_{0}$ is the free-space permittivity, we calculate the necessary Born effective charges ($Z_{\text{Lu/N}}^{*}=\pm3.48q$) to reproduce the experimental LuN LO($\Gamma$) frequency $587$~cm\textsuperscript{-1} ($72.8$~meV) using $\omega_{\text{LO}}^{2}=\omega_{\text{TO}}^{2}+(Z^{*})^{2}/(\Omega\varepsilon_{\infty}\mu)$ where $q$ is the effective charge, $\Omega$ is the volume of the primitive unit cell and $\mu=(m_{\text{Lu}}^{-1}+m_{\text{N}}^{-1})^{-1}$ is the reduced mass. The derived Born effective charge is similar to a recent measurement in the electronically-similar ScN, which found a value of $Z_{\text{Sc/N}}^{*}=\pm3.81q$~\cite{grumbel2024band}.

Using the $12\times12\times12$ $\mathbf{k}$ point grid we converged the force constants using a range of $\mathbf{q}$ point grids, the convergence test for which is shown in Figure~\ref{fig:LuN_convergence}(b). The real-space force constants converged using a $6\times6\times6$ $\mathbf{q}$ point grid and were then Fourier-transformed onto an arbitrary $\mathbf{q}$ grid to compute the phonon dispersion along a high-symmetry path, a generalized phonon density of states with a uniformly spaced $30\times30\times30$ $\mathbf{q}$ grid, and an evenly spaced grid of $\mathbf{q}$ points on the surface of a sphere folded back into the primitive Brillouin zone, visualized using Figure~\ref{fig:Q} for momentum transfers $1\leq|Q|\leq6$~{\AA}\textsuperscript{-1}. The sampling of points on the sphere was designed to mimic the momentum transfer probed using the inelastic neutron scattering experiment. Low momentum transfers sample a uniform sphere of points within the first Brillouin zone. As the momentum transfer is increased, the area of the sphere increases, and the sampling becomes more uniform due to the multiple folding of the surface of $\mathbf{q}$ points into the first Brillouin zone. The GDOS and $e$-GDOS from the two were scaled in contribution to the generalized phonon density of states by the NIST thermal neutron scattering cross sections of Lu ($6.53$~barn) and N ($11.51$~barn), and the Bose occupation factor at $4$~K~\cite{sears1992neutron}. The experimental broadening of the INS experiment is reflected in the GDOS and $e$-GDOS by broadening them by $0.9$~meV.

\subsection{Experimental methods}

The LuN powder was prepared using high energy ball milling of metallic lutetium (Lu) filings in a nitrogen glovebox~\cite{Kneisel2024}. The metallic Lu filings deform and are reduced in size, allowing new surfaces to be constantly exposed to nitrogen gas, eventually converting to LuN powders used for further experiments. There is only minimal sample contamination originating from Lu phases incorporated during sample preparation. These materials were kept in a nitrogen glove box except during XRD and Raman measurements using a Rigaku SmartLab x-ray diffractometer with a Cu source and Jobin Yvon LabRAM spectrometer using an argon ion laser, respectively, for which custom air-tight sample holders have been designed.

All neutron scattering measurements were conducted using Taipan, a thermal neutron TAS with o-40’-40’-o collimation with an incident pyrolytic graphite(002), PG(002), monochromator located at ANSTO~\cite{danilkin2009taipan,danilkin2012taipan,rule2018recent}. The measurements were conducted using a fixed final neutron energy of $14.87$~meV for suppressing contributions originating from $\lambda/2$ and $\lambda/3$ neutrons by using a $5$~cm PG transmission filter. The neutron diffraction measurements of the LuN sample ($4.69$~g) in an annular aluminium can were conducted at $300$ and $4$~K using a cryostat and a thermal neutron wavelength of $\lambda_{\text{n}}=2.345$~{\AA}. INS measurements were conducted in the energy loss configuration with small ($|\vec{Q}|=3.0$~{\AA\textsuperscript{-1}}) and large ($|\vec{Q}|=4.5$~{\AA\textsuperscript{-1}}) momentum transfer and energy transfer between $3$~meV and $55$~meV on our LuN powder at $4$~K. The uncertainties were taken as the square root of the neutron flux. The energy resolution for Taipan is $0.9$~meV at the elastic line~\cite{chen2026uncovering}. Due to geometric limitations with the PG monochromator and drastically increased integration times necessary with the Cu monochromator, no higher energy transfers were tested.

\bibliography{Ram}

\end{document}